\documentclass[twocolumn,showpacs,amsmath,amssymb,prl]{revtex4-1}

\usepackage[pdftex]{graphicx}
\usepackage[pdftex,bookmarks=true,bookmarksopen,bookmarksnumbered,
                colorlinks,
                linkcolor=blue,
                citecolor=blue]{hyperref}
\usepackage{dcolumn}
\usepackage{bm}

\begin{document}


\title{Observation of zero-point quantum fluctuations of a single-molecule magnet through the relaxation of its nuclear spin bath}

\author{A.~Morello$^{1,2}$, A.~Mill\'an$^3$, and L.J.~de Jongh$^1$}
\affiliation{%
$^1$Kamerlingh Onnes Laboratory, Leiden University, P.O. Box 9504,
2300RA Leiden, The Netherlands\\
$^2$ Centre for Quantum Computation and Communication Technology, School of Electrical
Engineering and Telecommunications, University of New South Wales,
Sydney NSW 2052, Australia\\
$^3$Instituto de Ciencia de Materiales de Arag\'on,
C.S.I.C. Universidad de Zaragoza, 50009 Zaragoza, Spain
}%

\date{\today}

\begin{abstract}
A single-molecule magnet placed in a magnetic field perpendicular to its anisotropy axis can be truncated to an effective two-level system, with easily tunable energy splitting. The quantum coherence of the molecular spin is largely determined by the dynamics of the surrounding nuclear spin bath. Here we report the measurement of the nuclear spin--lattice relaxation in a single crystal of the single-molecule magnet Mn$_{12}$-ac, at $T \approx 30$ mK in perpendicular fields $B_{\perp}$ up to 9 T. Although the molecular spin is in its ground state, we observe an increase of the nuclear relaxation rates by several orders of magnitude up to the highest $B_{\perp}$. This unique finding is a consequence of the zero-point quantum fluctuations of the Mn$_{12}$-ac spin, which allow it to efficiently transfer energy from the excited nuclear spin bath to the lattice. Our experiment highlights the importance of quantum fluctuations in the interaction between an `effective two-level system' and its surrounding spin bath.
\end{abstract}

\newcommand{\ket}[1] {\lvert #1 \rangle}


\maketitle

Quantum mechanical two-level systems (TLS) are the subject of vivid interest, motivated by their application in quantum information technology \cite{ladd10N}. In this context, the model of a `central spin + spin bath' \cite{prokof'ev00RPP} is of widespread fundamental interest for a wide range of natural or artificial TLS \cite{buluta11RPP}. Single-molecule magnets (SMMs) \cite{gatteschi03AC} constitute a prototypical example of mesoscopic quantum systems that, under suitable experimental conditions, can be treated as an effective TLS \cite{morello06PRL}. The `qubit levels' arise from the low-energy truncation of the larger Hilbert space of a high-spin (typical $S \sim 10$) molecule. In the presence of uniaxial magnetocrystalline anisotropy, the resulting qubit energy splitting $\hbar \omega_e$ has a very strong dependence on the external magnetic field $B_{\perp}$ applied perpendicular to the anisotropy axis \cite{korenblit78SPJETP}, $\hbar \omega_e \propto B_{\perp}^{2S}$, making the splitting easily tunable. Being stoichiometric and crystalline compounds, SMMs are exquisitely suited for the fundamental study of decoherence in mesoscopic systems, because the Hamiltonian of the qubit \emph{and} the environment (spin and phonon bath) is known in utmost detail. Indeed, a recent experiment found that the spin coherence of Fe$_8$ SMMs in large transverse field \cite{takahashi11N} is excellent agreement with the most accurate theories \cite{morello06PRL}.

\begin{figure}[tb] \center
\includegraphics[width=1\linewidth]{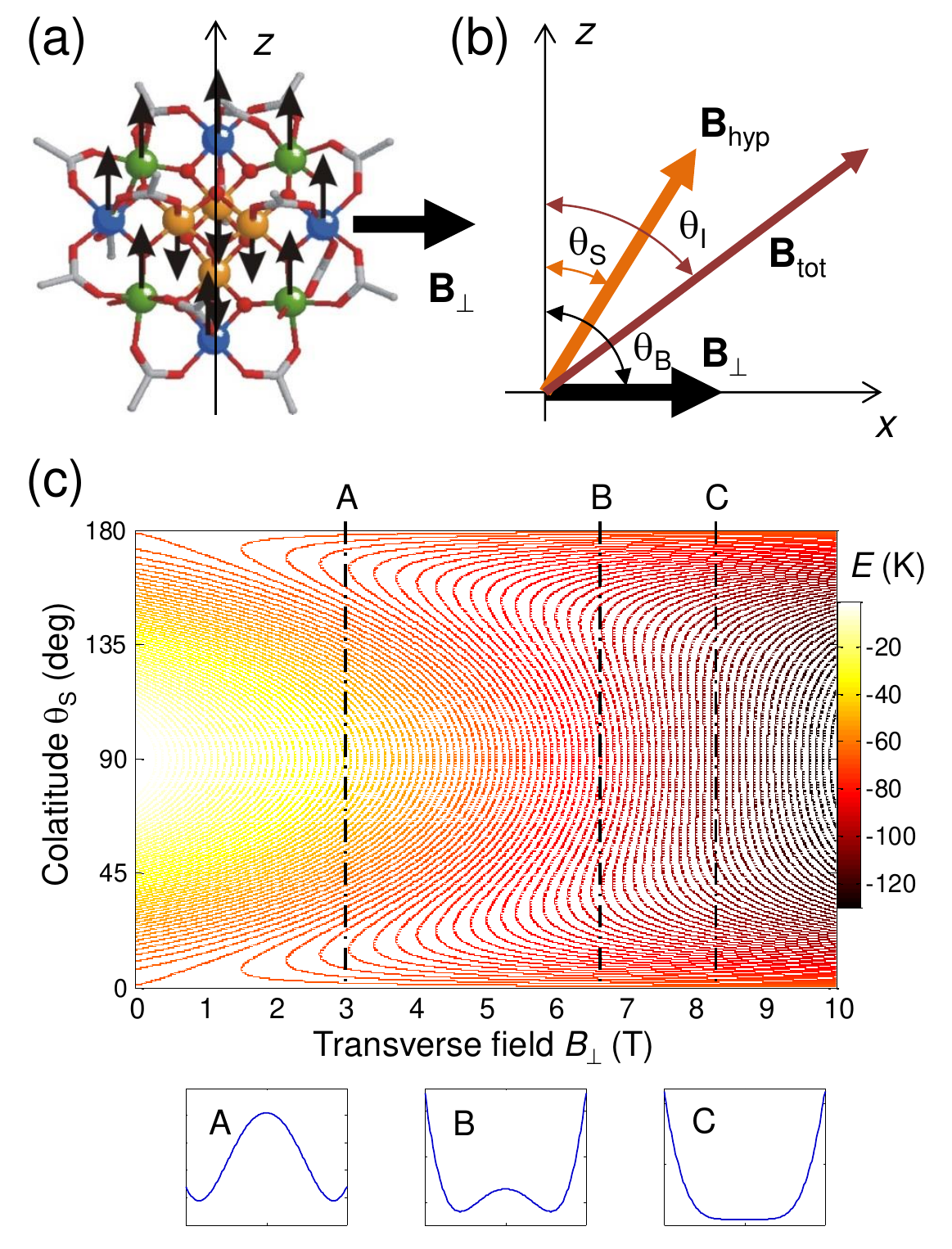}
\caption{ \label{fig_potential} (Color online) (a) Structure of the Mn$_{12}$-ac molecule. (b) Orientation of
the external ($\mathbf{B}_{\perp}$), hyperfine ($\mathbf{B}_{\rm
hyp}$) and total ($\mathbf{B}_{\rm tot}$) field on the Mn1 nuclei. (c) Evolution of the total energy $E$ (crystal field anisotropy + Zeeman term) of the Mn$_{12}$-ac spin as a function of transverse field and angle $\theta_{S}$. Representative energy profiles at $B_{\perp} = 3$~T (A), 6.65~T (B), 8.25~T (C) are shown underneath.
}
\end{figure}

Here we report a pulse-NMR study of the nuclear spin bath dynamics in a single crystal of the molecular magnet Mn$_{12}$-ac SMMs at ultra-low temperatures ($T \approx 30$~mK) in strong perpendicular magnetic fields ($B_{\perp}$ up to 9~T). This allows us to explore the regime where the molecular spin (qubit) splitting $\hbar \omega_e$ becomes larger than any other energy scale. We focus however on the \emph{bath} rather than the qubit, and find strong evidence that zero-point quantum fluctuations (ZPFs) \cite{clerk10RMP} of the molecular spin dominate the nuclear bath dynamics. We deduce this by measuring the transverse field dependence of the nuclear spin relaxation rate $1/T_{1n}(B_{\perp})$. In a material such as Mn$_{12}$-ac, nuclear spins do not have a channel for direct relaxation to the phonon bath -- their relaxation must be mediated by electron spin fluctuations. When $\hbar \omega_e \gg kT$, thermal fluctuations of the electron spin become exponentially suppressed and one expects $1/T_{1n}\rightarrow0$. Instead we find that $1/T_{1n}(B_{\perp})$ \emph{increases by five orders of magnitude} up to the highest field. Indeed, quantum mechanics predicts \cite{zvezdin98PRB} that, since a fully polarized state with  $\textbf{S}\parallel\textbf{B}_{\perp}$ is \emph{not} an eigenstate of the spin hamiltonian (unless $B_{\perp} \rightarrow \infty$), the central spin will exhibit quantum fluctuations down to $T=0$. This experiment thus represents an attractive physical implementation of the ideas discussed by Gavish \textit{et al.} \cite{gavish00PRB}, who argued that, although ZPFs cannot \textit{supply} energy, they can indeed (even at $T \approx 0$) \textit{absorb} energy when coupled to an activated system (here the nuclear spins, excited by NMR pulses), thereby de-exciting it.

The properties of the Mn$_{12}$-ac SMM are well known. We adopt the same electron spin Hamiltonian $\mathcal{H}_{S}$ and parameter values as in previous work \cite{morello07PRB,supp}:
 \begin{equation}
\mathcal{H}_S=DS_{z}^{2}+E(S_{x}^{2}-S_{y}^{2})+B_{4}S_{z}^{4}+C(S_{+}^{4}+S_{-}^{4})+\mu_{B}\mathbf{B}_{\perp}\cdot
\hat{g}\cdot \mathbf{S}. \label{hamiltonian}
\end{equation}
 We also add a term accounting for a dipolar field $B_{\rm dd}$ from neighboring molecules, of the form $\mathcal{H}_{\rm dd} = g_z \mu_B B_{\rm dd} S_z \approx 0.1$~K \cite{tupitsyn04CM}.
 The hyperfine interaction between nuclear and electron spins is $\mathcal{H}_{hyp}= - \mathbf{I} \cdot \hat{A} \cdot \mathbf{S}$ , with $\hat{A}$ the hyperfine tensor. The Mn$_{12}$-ac molecules in the crystal contain 4 Mn$^{4+}$ ions (with ionic spin $s=3/2$), giving rise to the NMR resonance labeled Mn1, and 8 Mn$^{3+}$ ions ($s=2$) [Fig.~\ref{fig_potential}(a)], the latter occupying two inequivalent sites which give rise to different hyperfine interactions \cite{kubo02PRB} and two separate resonance lines, Mn2 and Mn3. \textit{Intra}molecular magnetic exchange interactions between the ionic spins yield a net effective spin $S=10$ for the molecular cluster at low $T$. Below $\sim 10$~K only the lowest doublet of states $\lvert \mathcal{G} \rangle , \lvert \mathcal{E} \rangle$ is thermally occupied, justifying the truncation of the `giant spin' to an effective qubit Hamiltonian. When $B=0$, $\ket{\mathcal{G}}$ and $ \ket{\mathcal{E}}$ are symmetric and antisymmetric quantum superpositions of the $m_S = \pm 10$ projections of the molecular spin along the $z$-axis. Their energy splitting arises from the (weak) off-diagonal terms in $\mathcal{H}_{S}$, which introduce a tunnel coupling $\Delta_0$ ($\sim 10^{-10}$~K at $B=0$) between spin states at opposite side of the classical spin anisotropy barrier [Fig.~\ref{fig_potential}(c)]. In previous work \cite{morello04PRL} we showed that the nuclear spin bath can relax and thermalize via incoherent quantum tunneling of the central spin, which is itself \emph{driven} by the internal dynamics of the nuclear bath \cite{prokof'ev00RPP}, giving rise to non-trivial quantum relaxation effects \cite{wernsdorfer99PRL}. It was also argued that the most of the dynamics at $B=0$ arises from a minority of fast-relaxing molecules (FRMs) \cite{morello04PRL,wernsdorfer99EPL}.

\begin{figure}[bt] \center
\includegraphics[width=1\linewidth]{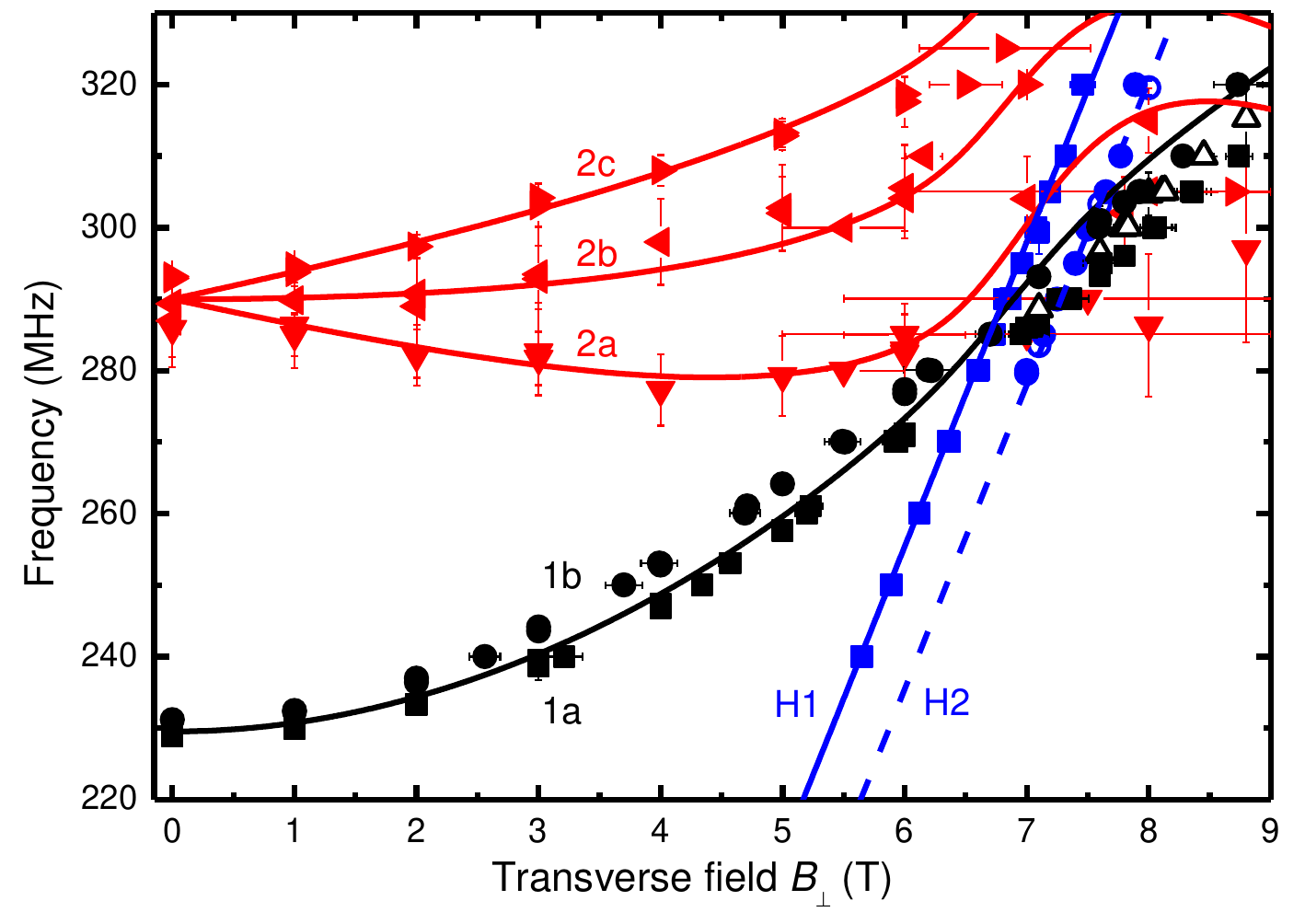}
\caption{ \label{fig_spectra} (Color online) Field dependence of the Mn1, Mn2 and $^{1}$H
NMR resonances (symbols) and FWHM (error bars, horizontal or vertical,
from swept-field and -frequency runs, resp.). Lines are
simulations discussed in supplementary material.
}
\end{figure}

In the present study we apply a strong perpendicular field $B_{\perp} \parallel x$, which causes $\ket{\mathcal{G}}$ and $\ket{\mathcal{E}}$ to contain amplitudes from all $m_S = -10~\ldots~+10$. They can still be written as a superposition of `classical' states $\ket{\mathcal{Z}_{\pm}}$ corresponding to the two total energy minima [Fig.~\ref{fig_potential}(c)], but the spin expectation values of $\ket{\mathcal{Z}_{\pm}}$ are now canted towards the $x$-axis \cite{furukawa03PRB}, forming an angle $\theta_S = \sin^{-1}\left(\langle S_x \rangle / S\right)$ with the $z$-axis [Fig.~\ref{fig_potential}(b)]. The hyperfine interaction $\mathcal{H}_{hyp}$ is commonly written in terms of an effective field $\textbf{B}_{\rm hyp}= - \hat{A} \cdot \textbf{S}/\hbar\gamma_{n}$. The total field at the nuclear site, $\textbf{B}_{\rm tot} = \textbf{B}_{\perp} + \textbf{B}_{\rm hyp}$, defines the quantization axis for the nuclear spin and yields the nuclear Larmor frequency $\omega_{n}=2\pi\nu_{n}=\gamma_{n} |\textbf{B}_{\rm tot}|$, with $\gamma_{n}$ the nuclear gyromagnetic ratio. We performed $^{55}$Mn ($I=\frac{5}{2}$) and proton ($I=\frac{1}{2}$) NMR experiments in the frequency range 220-330~MHz, in a cryogenic set-up described elsewhere \cite{morello05RSI,morello07PRB}. The Mn$_{12}$-ac single crystal was carefully oriented with its $z$-axis perpendicular to the magnetic field direction. Several attempts were made to fine-tune the crystal orientation, with no significant changes. Fig.~\ref{fig_spectra} shows the evolution of the NMR frequencies with $B_{\perp}$.

The Mn1 line is clearly split in two (1a and 1b in Fig.~\ref{fig_spectra}), as also observed in other Mn$_{12}$-ac single crystal studies in zero field \cite{harter05IC,morello07PRB}. Its field dependence is explained by a progressive canting of the total spin by $B_{\perp}$, and assuming $S=10$ at all fields \cite{furukawa03PRB}. The strong splitting of the Mn2 line was not observed in a previous experiment on aligned powder \cite{furukawa03PRB}, but is readily explained on basis of the presence of the anisotropic dipolar hyperfine field \cite{kubo02PRB,supp}. The Mn3 line falls mostly outside our measurement window and will not be discussed. Two proton lines enter the measurement window when $5\leq B_{\perp}\leq 8$~T. The main one (H1), responsible for $\approx90$\% of the estimated total proton intensity, is observed at the unshifted frequency $\gamma B_{\perp}/2\pi$, implying a (dipolar) hyperfine field $B_{\rm dip}\approx$~0.1~T or less. The second line (H2) is shifted by $-19$~MHz, corresponding to a $B_{\rm dip}\approx$~0.5 T, in agreement with earlier deuteron NMR data \cite{dolinsek98EPL}, taking into account the difference in nuclear moments. The calculated $^{55}$Mn spectra, shown as solid lines in Fig.~\ref{fig_spectra}, agree well with the data when assuming $\mathbf{B}_{\perp}$ makes an angle $\approx 90^{\circ} \pm 2^{\circ}$ with the $z$-axis.

\begin{figure}[bt] \center
\includegraphics[width=1\linewidth]{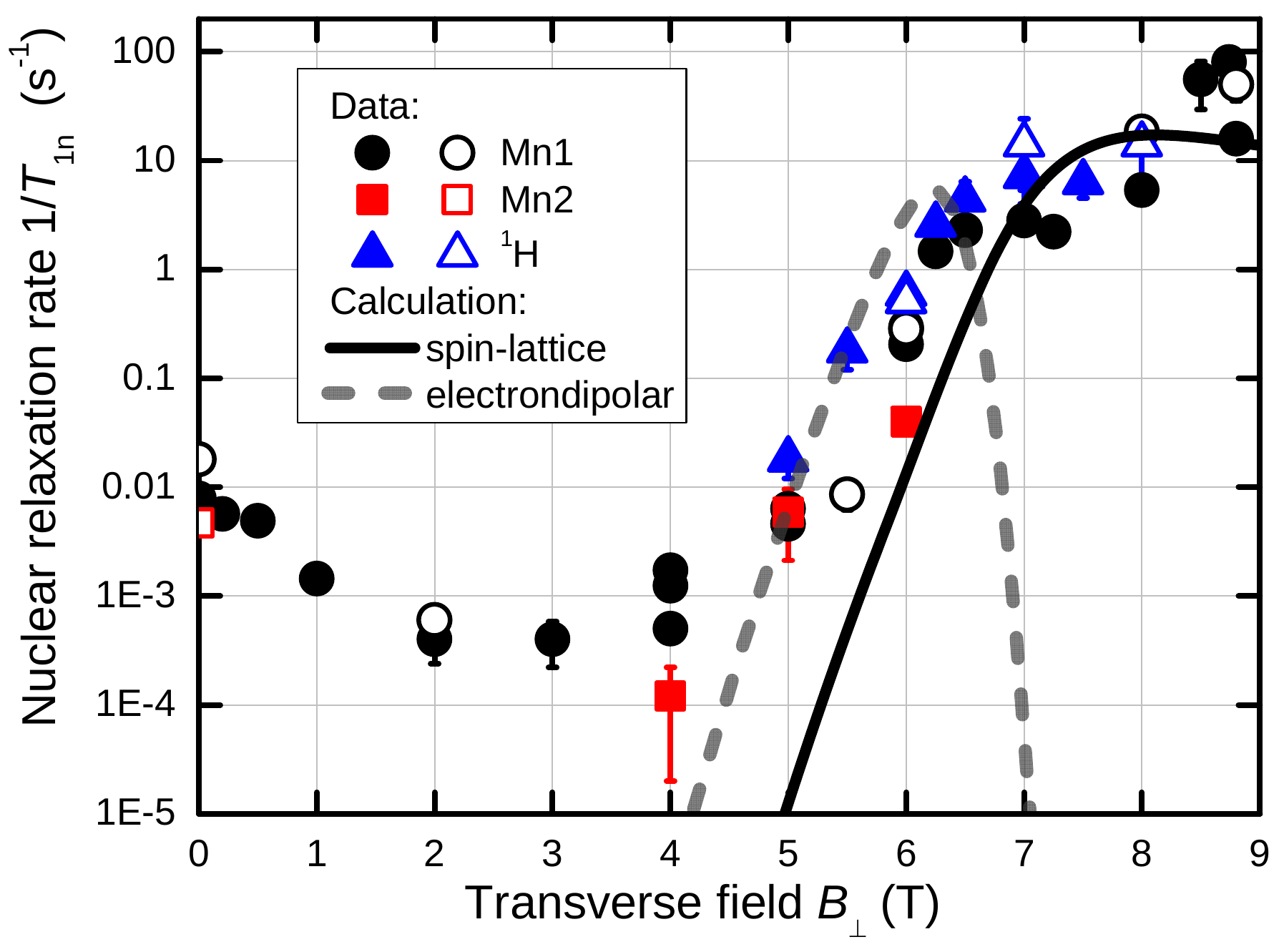}
\caption{ \label{fig_T1} (Color online) (a) Nuclear spin relaxation rates $1/T_{1n}$ versus $B_{\perp}$
for Mn1, Mn2 and H1. Open symbols are data taken at a slightly ($\sim 2^{\circ}$) different crystal orientation. Solid line: calculated $1/T_{1n}$ from direct spin-lattice relaxation. Dashed line: calculated $1/T_{1n}$ from two-step relaxation through electron dipolar reservoir.
}
\end{figure}

\begin{figure}[bt] \center
\includegraphics[width=1\linewidth]{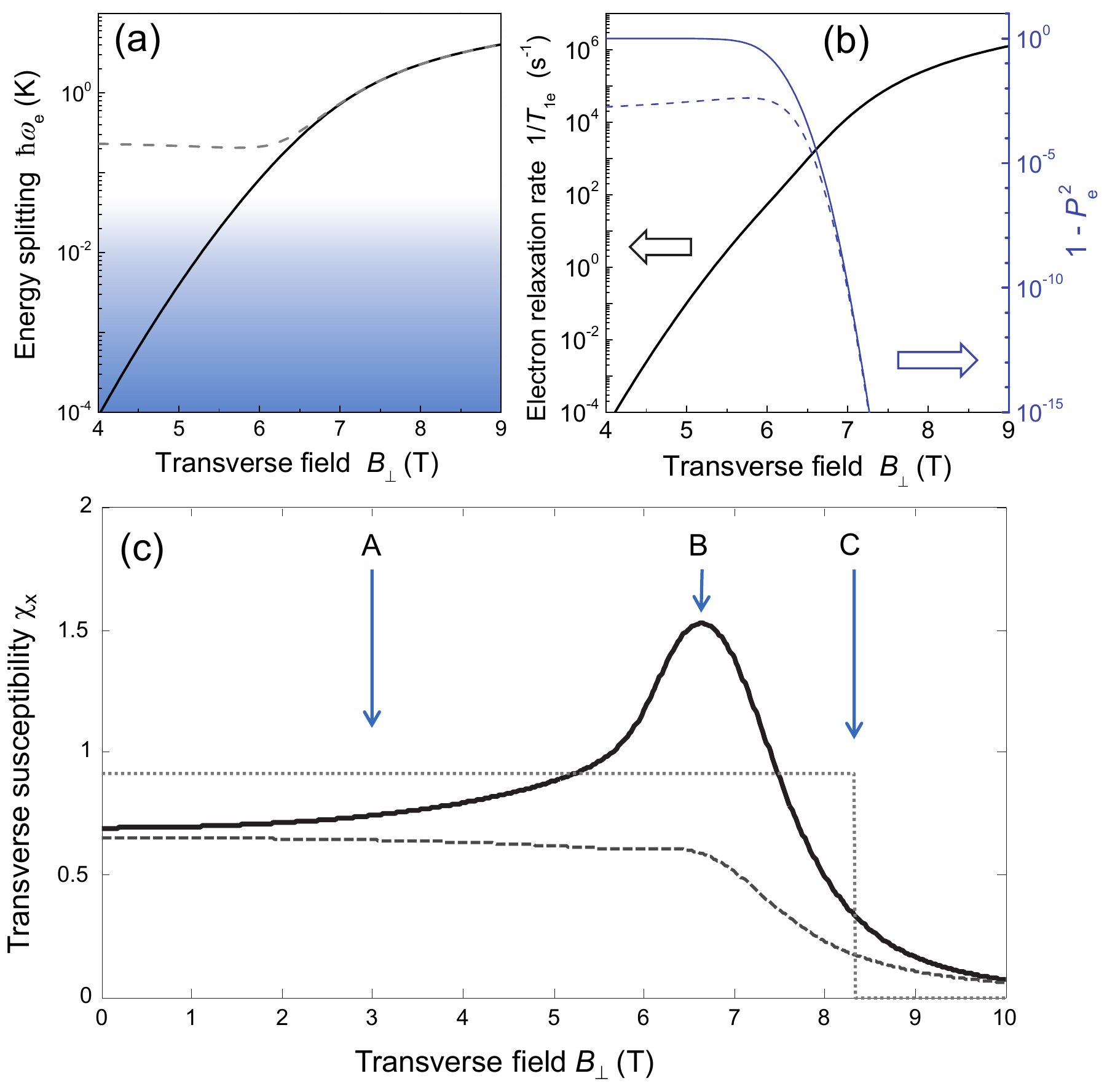}
\caption{ \label{fig_chi} (Color online) (a) Energy splitting $\hbar \omega_e$ of the lowest electron spin state of the Mn$_{12}$-ac molecule. Solid line: tunnel splitting $2\Delta_0$. Dashed line: splitting including a longitudinal dipolar bias field, $\hbar \omega_e = \sqrt{(g \mu_B B_{\rm dd})^2 + 2\Delta_0^2}$. For $B_{\perp} > 6.5$~T, $2\Delta_0$ exceeds both $g \mu_B B_{\rm dd}$ and $k_B T$ ($T \approx 30$~mK, shaded area). (b) Calculated electron spin relaxation rate $1/T_{1e}$ (left scale) and $(1-P_e^2)$ (right scale), where $P_e$ is the electron spin polarization. The dotted line includes the contribution of $\mathcal{H}_{\rm dd}$. (c) Mn$_{12}$-ac spin susceptibilities. Solid line: full quantum mechanical calculation of $\chi_x$. Dashed: spin fluctuations $\langle S_x^2 \rangle - \langle S_x \rangle^2$. Dotted: classical mean-field value. Arrows show the points corresponding to the anisotropy potentials A, B, C in Fig.~\ref{fig_potential}(c) as indicated.
}
\end{figure}

The longitudinal nuclear spin relaxation (NSR) rate, $1/T_{1n}$, was measured with inversion recovery sequences,
$\pi$-$t$-$\frac{\pi}{2}$-$\tau$-$\pi$ \cite{supp}. A compilation of the NSR rates as a function of $B_{\perp}$ for Mn1, Mn2 and H1 is given in Fig.~\ref{fig_T1}. We shall concentrate on the Mn1 line, which we were able to follow from $B_{\perp}=0$ to 9~T, but all lines behave similarly in the field range where they could be observed. For moderate $B_{\perp}$, $1/T_{1n}$ initially decreases, consistently with the interpretation that the $B \approx 0$ relaxation processes are driven by fast-relaxing molecules \cite{morello04PRL}, whose easy axis is misaligned with that of the crystal. However, above $\sim 5$~T a spectacular increase of a factor $10^{4}-10^{5}$ is seen in the $1/T_{1n}$ of all lines, all the way to 9~T. To appreciate the significance of this observation, let us recall the standard expression of the NSR rate for nuclear spins coupled to a paramagnetic electron spin \cite{abragam78RPP}:
\begin{equation}
\frac{1}{T_{1n}} \approx \frac{1}{T_{1e}} \left(\frac{A_{\perp}}{\omega_n}\right) (1-P_e^2), \label{waugh}
\end{equation}
where $A_{\perp}$ is the off-diagonal part of the hyperfine coupling, $P_e = \tanh \left(\hbar \omega_e / 2kT\right)$ is the electron spin polarization, and $1/T_{1e}$ is the electron spin-lattice relaxation rate \cite{morello06PRL}. For Mn$_{12}$-ac the latter is given by:
\begin{eqnarray}
\frac{1}{T_{1e}} \simeq \frac{4D^2|\langle \mathcal{E}|S_x S_z + S_z
S_x|\mathcal{G}\rangle|^2 (\hbar\omega_e)^3}{3 \pi \rho c_s^5
\hbar^4}\coth \left( \frac{\hbar \omega_e}{2k_{\rm B}T} \right),
\end{eqnarray}
 where $\rho=1.83\times10^{3}$ kg/m$^3$ is the density and $c_{s}\simeq 1.5\times10^{3}$~m/s is the sound velocity. In Fig.~\ref{fig_chi}(b) we plot the calculated values of $1/T_{1e}$ and $\left(1 - P_e^2\right)$ as a function of $B_{\perp}$. Although $1/T_{1e}$ does increase with field, $\left(1 - P_e^2\right)$ decreases much more dramatically for $B_{\perp} > 6.5$~T. Naively applying Eq.~(\ref{waugh}) to our system would lead to the (incorrect) prediction of astronomically long nuclear spin relaxation times for $B_{\perp} > 7$~T. Therefore we need to reconsider the problem of nuclear spin-lattice relaxation, taking proper account of the complex nature of the Mn$_{12}$-ac electronic spin $S$, and recognizing that the factor $(1-P_e^2)$ in Eq.~(\ref{waugh}) represents in fact the differential electronic susceptibility $\chi_{\alpha} = g \mu_B \partial \langle S_{\alpha}\rangle / \partial B_{\alpha}$, with $\alpha$ the direction of the applied field.

The nuclear magnetization is relaxed via random fluctuations $\delta B_{hyp}(t)=-(A_{\perp}/\hbar\gamma_{n})~\delta S(t)$ in the hyperfine fields, associated with the electron spin fluctuations $\delta S(t)$. The NSR rate is obtained in perturbation theory as the Fourier transform of the electron spin correlation functions, evaluated at the nuclear Larmor frequency $\omega_n$:
\begin{eqnarray}
\frac{1}{T_{1n}}=\frac{1}{2} \left(\frac{A_{\perp}}{\hbar}\right)^{2}\bar{F}_{\alpha}(\omega_n), \\
\bar{F}_{\alpha}(\omega)=\int \langle \{\delta S_{\alpha}(0) \delta S_{\alpha}(t)\} \rangle e^{-i\omega t}dt \ \ (\alpha=x,y,z)
\end{eqnarray}
 Here $\langle~\rangle$ denotes the thermal statistical average and $\{~\}$ the symmetrized spin operator product;  $\bar{F}(\omega)=(1/2)[F(\omega)+F(-\omega)]$ is the symmetrized quantum spectral density \cite{kubo66RPP,gavish00PRB,clerk10RMP} and positive/negative frequencies correspond to absorption/emission of energy. Detailed balance requires $F(\omega)= e^{\hbar\omega/k_{B}T} F(-\omega)$ and thus: $\bar{F}(\omega)=(1/2)\coth(\hbar\omega/2k_{B}T)[F(\omega)-F(-\omega)]$, leading to the low-$T$ quantum version of the fluctuation-dissipation theorem \cite{kubo66RPP,clerk10RMP}:
 \begin{equation}
 \bar{F}_{\alpha}(\omega)=\frac{\hbar}{g^{2}\mu_{B}^{2}}\coth\left(\frac{\hbar\omega}{2k_{B}T}\right)\chi_{\alpha}^{\prime \prime}(\omega),
 \end{equation}
 relating spin fluctuations to the imaginary component $\chi^{\prime \prime}$ of the complex magnetic susceptibility. Using the Debye frequency distribution and focusing on the \emph{transverse} susceptibility $\chi_x$, we write $\chi_x^{\prime \prime}(\omega) = \chi_x(0)\omega T_{1e} / (1+\omega^2 T_{1e}^2)$ and arrive at the final expression for the field-dependent NSR:
 \begin{equation}
 \frac{1}{T_{1n}(B_{\perp})} = \frac{\hbar}{2} \left(\frac{A_{\perp}}{g \mu_B}\right)^2 \coth \left( \frac{\hbar \omega_n}{2k_B T}\right) \chi_x \frac{\omega_n T_{1e}}{1+\omega_n^2 T_{1e}^2}, \label{T1n}
 \end{equation}
where $A_{\perp}$, $\omega_n$, $T_{1e}$ and $\chi_x$ are all functions of $B_{\perp}$. From the Mn$_{12}$-ac Hamiltonian (\ref{hamiltonian}), we calculate numerically $\chi_x(B_x) = g \mu_B \partial \langle S_x \rangle / \partial B_x$ [solid line in Fig.~\ref{fig_chi}(c)] and use it in Eq.~(\ref{T1n}) to obtain the NSR shown as the solid line in Fig.~\ref{fig_T1}. The calculation correctly predicts the strong increase in $1/T_{1n}$ up to the highest $B_{\perp}$.

The situation most often found in the literature has $\alpha=z$, in which case the electron Zeeman term $g\mu_{B}B_{z}S_z$ commutes with the strong uniaxial anisotropy term $DS_{z}^{2}$, $\hbar\omega_{e}$ is just the electron Zeeman splitting, and the susceptibility $\chi_{z}$ becomes the factor $(1-P_{e}^{2}$ in Eq.~(\ref{waugh}), which vanishes in high field as $\exp \left(\hbar \omega_e/k_B T\right)$. Our experiment is unique in that we have $\alpha=x$ and thus $g\mu_{B}B_{x}S_x$ does not commute with $DS_{z}^{2}$. The electronic splitting $\hbar\omega_{e}$ is now the quantum tunneling splitting, and the transverse susceptibility $\chi_{x}$ [Fig.~\ref{fig_chi}(c)] remains finite up to $B_{\perp} \rightarrow \infty$ even at $T=0$.

To further clarify the special role of ZPFs in our system, we recall that the susceptibility $\chi_x(0)$ can also be obtained from the static isothermal magnetization (per molecule) $\langle M_x\rangle=\sum_{i}(-\partial E_{i}/\partial B_x)e^{-\beta E_i}/\sum_{i}e^{-\beta E_i}$ ($\beta=1/k_{B}T$, and $E_{i}$ are the electronic energy levels) as:
 \begin{equation}
 \chi_x(0) = \beta\langle M_x^{2}\rangle-\beta\langle M_x\rangle^{2}+\langle-2\partial E_{i}/\partial B_x^{2}\rangle. \label{chi_static}
 \end{equation}
 The term $\langle-2\partial E_{i}/\partial B_x^{2}\rangle$ is the Van Vleck susceptibility, and describes a change in magnetization which does not originate from a change in thermal population of the eigenstates. Although often neglected, it plays a crucial role in our case since it persists even to $T$=0, and is responsible for the pronounced increase in  $\chi_x$ around $B_{\perp} \approx 6.65$~T, where the most drastic restructuring of the energy spectrum takes place \cite{zaslavskii84}. The susceptibility peak signals the onset of strong ZPF of the Mn$_{12}$-ac in the shallow double-well potential \cite{zvezdin98PRB}. The classical mean-field susceptibility $\chi_x(B_{\perp}) = (g \mu_B)^2 / 2D$ [Fig.~\ref{fig_chi}(c), dotted line] goes to zero for $B_{\perp} >2DS/g \mu_B$ ($= 8.32$~T in Mn$_{12}$-ac) and is clearly insufficient to explain the finite NSR up to 9 T. Neglecting the Van Vleck term from (\ref{chi_static}) is equivalent to considering only fluctuations of the form $\langle S_x^2 \rangle - \langle S_x \rangle^2$ [Fig.~\ref{fig_chi}(c), dashed line].


 An additional relaxation mechanism can arise due to the dipole-dipole coupling between molecular spins, as recently discussed in the case of the isotropic Mn$_6$ molecule \cite{morello06PRB}. In this process, nuclear energy is first shared very rapidly with the electron--dipolar (ED) reservoir, then the combined ED + nuclear--Zeeman (NZ) system relaxes to the lattice at the rate:
 \begin{equation}
 \frac{1}{T_{1n,e}^{\star}}= \frac{1}{T_{1e}} \frac{C_{ED}}{C_{NZ}+C_{ED}} \label{T1nstar}
 \end{equation}
where $C_{\rm ED}$ and $C_{\rm NZ}$ are the (field-dependent) specific heats of the ED and NZ reservoirs. We calculate $C_{\rm NZ}$ for the $^{55}$Mn as in Eq. 3 of Ref.~\onlinecite{morello06PRB}, and take for $C_{\rm ED}$ the Schottky curve for a two-level system. The resulting $1/T_{1n}^{\star}$ is plotted as gray dashed line in Fig.~\ref{fig_T1}. It drops markedly for $B_{\perp} > 6.5$~T due to the exponential decrease of  $C_{\rm ED}/R\approx(\hbar\omega_{e}/k_{B}T)^{2}e^{-\hbar\omega_{e}/k_{B}T}$ for $\hbar \omega_e \gg k_B T$, but it gives a significant contribution at lower fields. Taken together, the calculated values of the NSR from Eqs.~(\ref{T1n}) and (\ref{T1nstar}) are in remarkable qualitative and even quantitative agreement with the Mn1 data for $B_{\perp}\gtrsim$~5 T. We stress that no free fitting parameters were used at any point.

The -- perhaps surprising -- similarity between the Mn1, Mn2 and H1 relaxation rates (in the field range where data is available on all of them) can be related to the similar order of magnitude of the non-diagonal term $(A_{\perp}/\hbar\omega_{n})^{2}$ in Eq.~(\ref{T1n}). The value of $A_{\perp}$ depends on the angle between $B_{\rm hyp}$ and $B_{\rm tot}$ \cite{furukawa03PRB}, $A_{\perp} = A |\sin(\theta_I - \theta_S)|$ [see Fig.~\ref{fig_potential}(b)], as well as on the presence of a non-diagonal component in the hyperfine coupling tensor, more pronounced in Mn2 and H1.

In conclusion, we have shown that the relaxation of the nuclear spin bath in Mn$_{12}$-ac, at very low temperature, accelerates spectacularly when a large transverse field greatly enhances the zero-point quantum fluctuation of the molecular spin, whereas the electronic susceptibility remains finite up to highest applied field. The associated zero-point quantum fluctuations of the molecular spin provide a dynamics persisting even down to $T$=0 and thus a very effective channel for de-excitation of the nuclear spin bath, even when the Mn$_{12}$-ac spin can be considered as an effective two-level system in its ground state with a large gap to the first excited state. In addition to providing an appealing test bed for recent theories \cite{gavish00PRB}, our experiment highlights a profound difference between `true' $S = 1/2$ spin qubits \cite{koppens06N,pla12N}, and effective TLSs that arise from the low-energy truncation of more complex systems.

This work has been part of the research program of the ``Stichting
FOM''. We thank D. Bono for performing part of the experiments described. We acknowledge illuminating discussions with Y. Imry and R.S. Schoelkopf.


\end{document}